\documentclass[12pt,preprint]{aastex}

\shorttitle{Observations of Doppler Shift Oscillations}
\shortauthors{Mariska}

\begin{document}

\title{Observations of Solar Flare Doppler Shift Oscillations
  with the Bragg Crystal Spectrometer on {\it Yohkoh}}
\author{John T. Mariska}
\affil{E.\ O.\ Hulburt Center for Space Research, Code 7673,
Naval Research Laboratory, Washington, DC 20375}
\email{mariska@nrl.navy.mil}

\begin{abstract}
Oscillations in solar coronal loops appear to be a common
phenomenon.  Transverse and longitudinal oscillations have been
observed with both the {\it Transition Region and Coronal
Explorer} and Extreme Ultraviolet Imaging Telescope imaging
experiments.  Damped Doppler shift oscillations have been
observed in emission lines from ions formed at flare temperatures
with the Solar Ultraviolet Measurements of Emitted Radiation
Spectrometer.  These observations provide valuable diagnostic
information on coronal conditions and may help refine our
understanding of coronal heating mechanisms.  I have initiated a
study of the time dependence of Doppler shifts measured during
flares with the Bragg Crystal Spectrometer (BCS) on {\it Yohkoh}.
This Letter reports the detection of oscillatory behavior in
Doppler shifts measured as a function of time in the emission
lines of \ion{S}{15} and \ion{Ca}{19}.  For some flares, both
lines exhibit damped Doppler shift oscillations with amplitudes
of a few km~s$^{-1}$ and periods and decay times of a few
minutes.  The observations appear to be consistent with
transverse oscillations.  Because the BCS observed continuously
for almost an entire solar cycle, it provides numerous flare data
sets, which should permit an excellent characterization of the
average properties of the oscillations.
\end{abstract}


\section{INTRODUCTION}

Sequences of images made with the {\it Transition Region and
Coronal Explorer} ({\it TRACE}) show that a small subset of
large-scale loops appear to respond to nearby flares by
exhibiting damped transverse oscillations
\citep[e.g.,][]{Aschwanden99,Schrijver02}.  Damped Doppler shift
oscillations have also been detected over active regions with the
Solar Ultraviolet Measurements of Emitted Radiation (SUMER)
experiment on the {\it Solar and Heliospheric Observatory} ({\it
SOHO}) \citep[e.g.,][]{Kliem02,Wang03}.  These Doppler shifts are
primarily observed in emission lines from ions formed at
temperatures greater than 6 MK, and thus are probably associated
with flare-like events.

The study of these oscillations is fast developing into a new
technology for diagnosing the the physical conditions in the
loops and, hence, placing further constraints on possible coronal
heating mechanisms \citep[e.g.,][]{Roberts02,Roberts03}.

Detection of Doppler shifts in high-temperature emission lines
with SUMER suggests that they might also be detectable with
instruments capable of measuring Doppler shifts in soft X-ray
lines, such as the Bragg Crystal Spectrometer (BCS) on the {\it
Yohkoh} spacecraft.  \citet{Seely84} have in fact detected some
evidence for oscillations with a period of about 10 minutes in
the position of the \ion{Ca}{19} resonance line during the decay
of a limb flare observed on 1980 November 13 with the SOLFLEX
crystal spectrometer on the {\it P78-1} satellite.

In this Letter I show that damped Doppler shift oscillations are
present in some of the flares observed with the BCS.

\section{BCS OBSERVATIONS}

The {\it Yohkoh} BCS has been described in detail by
\citet{Culhane91}.  Four bent crystals cover narrow wavelength
ranges centered on emission lines of \ion{Fe}{26}, \ion{Fe}{25},
\ion{Ca}{19}, and \ion{S}{15}.  Radiation diffracted by each
crystal strikes a one-dimensional position-sensitive proportional
counter, with the \ion{Fe}{26} and \ion{Fe}{25} crystals sharing
one detector and the \ion{Ca}{19} and \ion{S}{15} crystals
sharing a second detector.  All four detectors view the entire
Sun.  For this study I use data from the \ion{Ca}{19} and
\ion{S}{15} channels.  Examples of the appearance of the spectra
in these two channels are available in, for example,
\citet{Doschek92}.

To search for Doppler shift oscillations in the BCS data, I have
initiated an examination of the temporal behavior of the physical
characteristics of a large number of the flares observed by the
BCS over its 10-year lifetime.  Fitting the BCS data using
synthetic spectra yields estimates of the temperature, emission
measure, nonthermal broadening, and Doppler shift of the emitting
plasma as a function of time.  The best fit parameters are
determined by computing synthetic spectra for an isothermal
plasma and then varying the fitting parameters to minimize the
$\chi^2$ statistic.

In most cases, the BCS obtains spectra in all four wavelength
channels every 3~s.  The \ion{Fe}{26} channel rarely contains
useful data, and obtaining Doppler shifts from the \ion{Fe}{25}
data is made difficult by the complexity of the spectra in that
channel.  Thus, this study is restricted to data from the
\ion{S}{15} and \ion{Ca}{19}.  Early or late in a flare, the
instrument count rates are small, resulting in spectra that are
difficult to fit.  For this survey, I have therefore summed
individual spectra until a minimum of 10,000 counts are
accumulated in the \ion{S}{15} and \ion{Ca}{19} channels.  This
generally results in excellent fits to the data and assures that
each data point in the resulting time series of physical
parameters will have roughly the same statistical significance.
It does, however, mean that the data points in the two derived
time series for each flare will not generally be at identical
times.

I have analysed data for more than 100 flares observed by the
BCS.  Most of the data examined to date cover the first year of
the mission.  In addition I have examined flares listed in
\citet{Schrijver02} for which there are useful BCS data.  A
significant number of the flares observed (roughly 25\%) show
evidence for oscillatory behavior.  In the following section, I
show the results for one event that exhibited evidence of damped
Doppler shift oscillations.

\section{RESULTS}

Figure~\ref{fig:bcs_data} shows the temporal behavior of the
count rates and the physical parameters resulting from fits to
the individual spectra for the \ion{S}{15} and \ion{Ca}{19}
wavelength channels in one flare that exhibits damped oscillatory
behavior.  This C2.5 flare took place at S16W88 on 1992 October
2, and is one of the nonocculted limb flares in the survey of
occulted and nonocculted limb flares performed by
\citet{mariska99}.  Nonocculted flares in that study refers to
events for which images obtained with the {\it Yohkoh} Soft X-Ray
Telescope (SXT) showed that the footpoints of the flaring loops
were not occulted by the solar limb.  At the peak of the flare,
the region of strong emission in the SXT images occupies only a
few pixels above the solar limb, suggesting that the hot plasma
detected with the BCS is confined to a relatively small volume.

While the total count rate, temperature, emission measure, and
nonthermal velocity show generally smooth behavior as a function
of time, the Doppler shift shows evidence of oscillatory behavior
in both channels.  In particular, the Doppler shift in the
\ion{S}{15} channel appears to show evidence for a damped
oscillation.

To determine the characteristics of the oscillations observed in
Figure~\ref{fig:bcs_data}, I follow the example of earlier work
on {\it TRACE} oscillations \citep[e.g.,][]{Aschwanden02} and fit
the observations with a combination of a damped sine wave and a
polynomial background.  For each channel I define a function of
the form
\begin{equation}
v(t) = A_0 \sin(\omega t + \phi)\exp(-\lambda t) + B(t)\, ,
\label{eq:sine_wave}
\end{equation}
where
\begin{equation}
B(t) = b_0 + b_1 t + b_2 t^2 + b_3 t^3 + \cdots
\end{equation}
is the trend in the background data.  Starting from an initial
guess for the fit parameters, the data in each channel were fit
to Equation~(\ref{eq:sine_wave}) using Levenberg-Marquardt
least-squares minimization \citep[e.g.,][]{Press92}.  For all the
flares studied the number of terms needed for the background
expression was generally one or two, although occasionally three
terms produced a better fit.

\begin{figure*}
\plotone{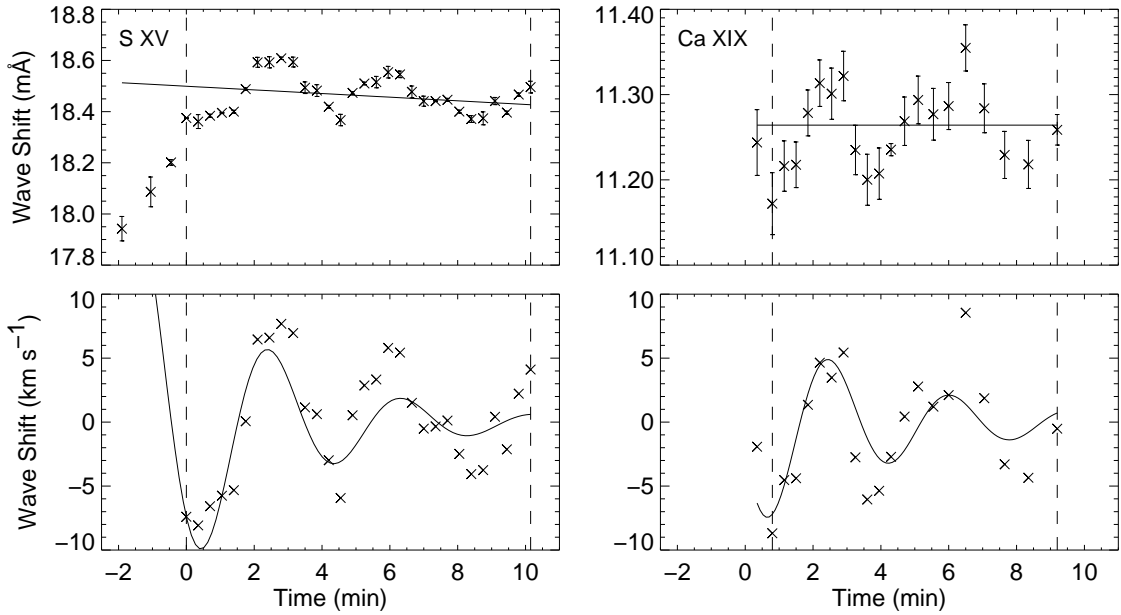}
\caption{Averaged BCS Doppler shift data and decaying sin wave
  fits for the \ion{S}{15} and \ion{Ca}{19} observations for the
  1992 October 12 flare.  The zero value for the shifts plotted
  in the top panels is based on an arbitrary reference system for
  the BCS in which a flare near the solar Equator with no Doppler
  shift would have a value of 0~m\AA.  Time is measured from
  21:50:44 UT.}
\label{fig:bcs_fits}
\end{figure*}

The Doppler shift data shown in Figure~\ref{fig:bcs_data} are
sufficiently noisy that it is difficult to fit them with
Equation~(\ref{eq:sine_wave}).  Since it is clear from the figure
that the period of the sine wave is on the order of a few
minutes, I have accumulated the BCS data for longer intervals by
requiring that the minimum number of counts in a channel exceed
10,000 and that the accumulation time be at least 20~s.  The top
panels in Figure~\ref{fig:bcs_fits} show the Doppler shift data
for the two channels accumulated in this manner.  Also plotted on
the top panels is the best fit background trend.  For the
\ion{S}{15} spectra, a two-term expression provided the best fit,
while for \ion{Ca}{19}, one term was sufficient.

The bottom two panels of Figure~\ref{fig:bcs_fits} show the
background-subtracted data along with the best-fit damped sine
wave.  For both channels, only the regions between the vertical
dashed lines were used in the fitting process.  Even with the
longer accumulation time the data are still noisy, especially
those from the \ion{Ca}{19} channel.  The fits to the data,
however, do appear to be reasonable.  Reduced $\chi^2$ values are
5.8 and 1.2 for the \ion{S}{15} and \ion{Ca}{19} channels,
respectively.

Table~\ref{tab:results} summarizes the fitting results.  While
the origin of time for the two channels was tied to the first
fitted spectrum, the values in the table for \ion{Ca}{19} have
been adjusted to correspond to the same initial time used for the
\ion{S}{15} fits.  The errors listed in the table correspond to
the diagonal elements in the covariance matrix for the fits.
Examination of the results in the table shows that both BCS
channels show evidence for the same oscillatory behavior in the
Doppler shift.  This was generally the case with the other flares
that showed damped oscillatory behavior.

\section{DISCUSSION AND CONCLUSIONS}

\begin{figure*}
\plotone{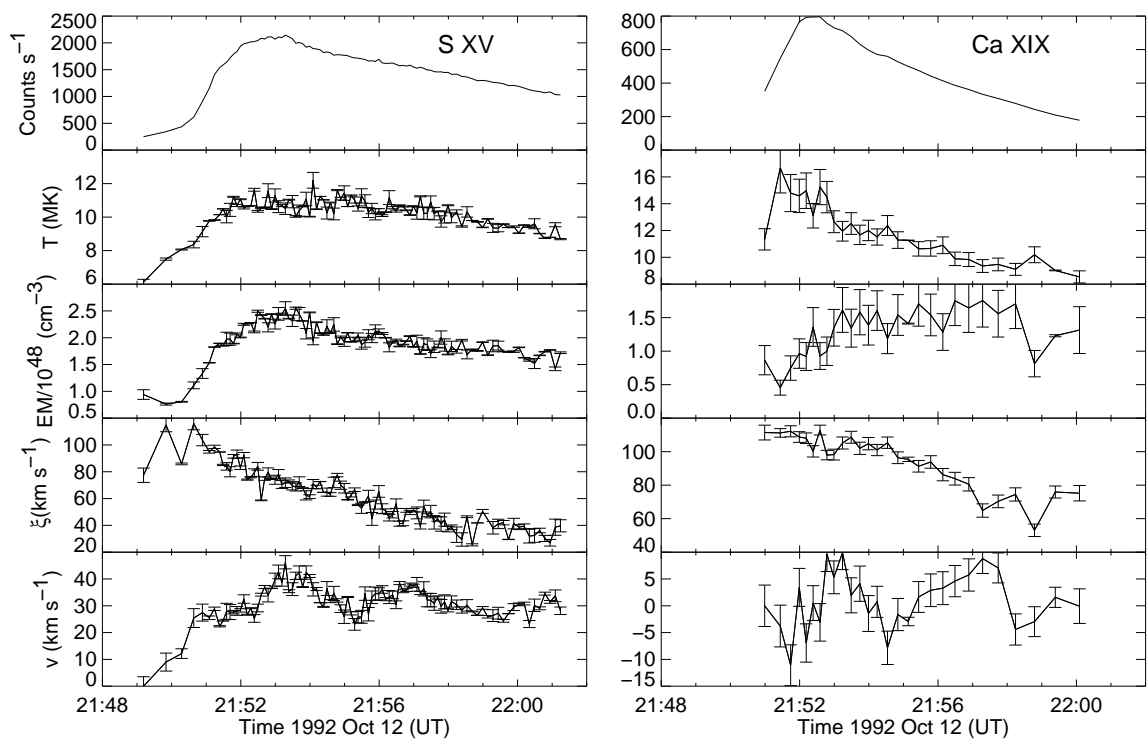}
\caption{Temporal behavior of the temperature, emission measure,
  nonthermal broadening velocity, and Doppler shift derived from
  the BCS \ion{S}{15} and \ion{Ca}{19} observations for the 1992
  October 12 flare.  The zero value for the Doppler shift
  velocities has been set to the wavelength shift of the first
  fitted spectrum.}
\label{fig:bcs_data}
\end{figure*}

Figure~\ref{fig:bcs_data} shows that for the first several
minutes of the flare the BCS \ion{S}{15} and \ion{Ca}{19}
channels show different plasma temperatures.  The comparable
oscillation parameters in Table~\ref{tab:results} thus indicate
that the flaring structures at the two temperatures are closely
related.  This suggests that the oscillations are not restricted
to just one isolated loop.

The measured values for the periods (3.93 and 3.57~min) and decay
times (3.52 and 4.26~min) fall near the low side of the range of
values listed for the 26 oscillating loops observed with {\it
TRACE} by \citet{Aschwanden02}.  Following \citet{Wang03}, we
define the maximum displacement amplitude for each fit as
\begin{equation}
A = A_0/(\omega^2 + \lambda^2)^{1/2} \, .
\end{equation}
This yields displacements of 416 and 295~km for \ion{S}{15} and
\ion{Ca}{19}, respectively.  These values are less than 1~arcsec.
Thus, if the oscillations are due to a position change in the
magnetic structure confining the flaring plasma, the change would
be invisible at the roughly 3~arcsec spatial resolution of the
SXT.  This inferred excursion also falls near the low side of the
range listed by \citet{Aschwanden02}.  All the above values fall
below the ranges listed by \citet{Wang03} for Doppler shift
oscillations observed with SUMER.

The measured values for the oscillations shown in the the table
and figures are typical of those determined for the other flares
with well-defined oscillations.  Those flares tend to be at or
near the solar limb, which, since BCS measures only line-of-sight
Doppler shifts is where one would expect to most clearly see
oscillatory behavior.  Many other flares in the sample show
evidence for significant Doppler shift fluctuations, but the data
are not easily fitted with a model such as that used in
Equation~(\ref{eq:sine_wave}).  This would suggest that there are
many cases where different structures in the flaring plasma are
moving in an uncoordinated fashion, leading to a complex signal
in the BCS, which observes all the flaring plasma.

Since the BCS has no imaging capability and integrates over all
the plasma contributing photons to the emission observed in each
wavelength channel, it is not possible to determine from the
Doppler shift data alone whether we are observing transverse or
longitudinal oscillations.  The periods and decay times measured
for the event shown in this Letter and those measured for other
events in the study are more typical of those seen in the {\it
TRACE} observations \citep{Aschwanden02}---suggesting that the
BCS observations are detecting transverse oscillations.  On the
other hand, the ratio of the decay time to the period is roughly
0.9, more typical of the results obtained by \citet{Wang03}.

Unfortunately, none of the flare-associated events listed in
\citep{Wang03} had useful BCS data.  A number of the events
analyzed by \citep{Aschwanden02} have useful BCS observations,
and those events are included in a more comprehensive analysis of
the BCS Doppler shift observations, which is currently in
progress \citep{Mariska05}.  The results of that analysis should
provide a firmer estimate of the range of oscillation properties
seen in the BCS data and help determine whether they are
transverse or longitudinal oscillations.

\acknowledgments
I thank G.~A. Doschek, H. Hudson, and H.~P. Warren for their
comments on the manuscript.  This research was supported by
ONR/NRL 6.1 basic research funds.

\clearpage

\begin{deluxetable}{lcc}
\tablecaption{Fitting Results\label{tab:results}}
\tablewidth{0pt}
\tablehead{
\colhead{Parameter} & \colhead{\ion{S}{15}} &
\colhead{\ion{Ca}{19}}}
\startdata
$A_0$ (km s$^{-1}$) & $11.3 \pm 1.9$ & $8.76 \pm 2.78$ \\
$\omega$ (rad min$^{-1}$) & $1.60 \pm 0.06$ & $1.76 \pm 0.17$ \\
$\phi$ (rad) & $-2.43 \pm 0.05$ & $-2.86 \pm 0.41$ \\
$\lambda$ (min$^{-1}$) & $0.284 \pm 0.090$ & $ 0.235 \pm 0.130$ \\
\enddata
\end{deluxetable}

\clearpage

\clearpage

\begin{thebibliography}{}

\bibitem[Aschwanden et al.(2002)]{Aschwanden02} Aschwanden,
  M. J., DePontieu, B., Schrijver, C. J., \& Title, A. M. 2002,
  \solphys, 206, 99

\bibitem[Aschwanden et al.(1999)]{Aschwanden99} Aschwanden,
  M. J., Fletcher, L., Schrijver, C. J., \& Alexander, D. 1999,
  \apj, 520, 880

\bibitem[Culhane et al.(1991)]{Culhane91} Culhane, J.~L.~et al.\
  1991, \solphys, 136, 89

\bibitem[Doschek et al.(1992)]{Doschek92} Doschek, G.~A.~et al.\
  1992, PASJ, 44, L95.

\bibitem[Kliem, et al.(2002)]{Kliem02} Kliem, B., Dammasch,
  I.~E., Curdt, W., \& Wilhelm, K.\ 2002, \apjl, 568, L61

\bibitem[Mariska(2005)]{Mariska05} Mariska, J.~T. 2005, \apj, in
  preparation

\bibitem[Mariska \& McTiernan(1999)]{mariska99} Mariska, J. T.,
  \& McTiernan, J. M. 1999, \apj, 514, 484

\bibitem[Press et al.(1992)]{Press92} Press, W. H., Teukolsky,
  S. A., Vetterling, W. T., \& Flannery, B. P. Numerical Recipes
  in C (2d ed.; Cambridge: Cambridge Univ. Press)

\bibitem[Roberts(2002)]{Roberts02} Roberts, B. 2002, \solphys,
  193, 139

\bibitem[Roberts(2003)]{Roberts03} Roberts, B. 2003, in SOHO13,
  Waves and Small Scale Transient Events in the Solar Atmosphere:
  A Joint View from SOHO and TRACE, ESA SP-547

\bibitem[Schrijver et al.(2002)]{Schrijver02} Schrijver, C. J.,
  Aschwanden, M. J, \& Title, A. M. 2002, \solphys, 206, 69

\bibitem[Seely \& Feldman(1984)]{Seely84} Seely, J. F., \&
  Feldman, U. 1984, \apj, 280, L59

\bibitem[Wang et al.(2003)]{Wang03} Wang, T.~J., Solanki, S.~K.,
  Curdt, W., Innes, D.~E., Dammasch, I.~E., \& Kliem, B.\ 2003,
  \aap, 406, 1105

\end{thebibliography}
\end{document}